\documentclass[preprint,12pt]{aastex}

\newcommand{\zabs}{$z_{\rm abs}$}

\usepackage{graphics,graphicx}
\usepackage{epsf,wrapfig}
\usepackage{color}
\usepackage{sidecap}

\begin{document}

\title{Quasar Absorption Lines in the Far Ultraviolet: An Untapped Gold Mine for Galaxy Evolution Studies}

\author{Todd M. Tripp [tripp@astro.umass.edu, (413)-545-3070]}

\affil{Department of Astronomy, University of Massachusetts-Amherst, Amherst, MA 01003}

\begin{abstract}
Most of the baryons are exceedingly difficulty to observe, at all
epochs. Theoretically, we expect that the majority of the baryonic
matter is located in low-density, highly ionized gaseous envelopes of
galaxies -- the ``circumgalactic medium'' -- and in the highly ionized
intergalactic medium. Interactions with the CGM and IGM are thought to
play crucial roles in galaxy evolution through accretion, which
provides the necessary fuel to sustain on-going star formation, and
through feedback-driven outflows and dynamical gas-stripping
processes, which truncate and regulate star formation as required in
various contexts (e.g., low-mass vs. high-mass galaxies; cluster
vs. field).  Due to the low density and highly ionized condition of
these gases, quasar absorption lines in the rest-frame ultraviolet and
X-ray regimes provide the most efficient observational probes of the
CGM and IGM, but ultraviolet spectrographs offer vastly higher
spectral resolution and sensitivity than X-ray instruments, and there
are many more suitable targets in the UV, which enables carefully
designed studies of samples of particular classes of objects.  This
white paper emphasizes the potential of QSO absorption lines in the
rest-frame far/extreme UV at $500 \lesssim \lambda _{\rm rest}
\lesssim 2000$ \AA . In this wavelength range, species such as
Ne~VIII, Na IX, and Mg X can be detected, providing diagnostics of gas
with temperatures $\gg 10^{6}$ K, as well as banks of adjacent ions
such as O~I, O~II, O~III, O~IV, O~V, and O~VI (and similarly N~I --
N~V; S~II -- S~VI; Ne~II -- Ne~VIII, etc.), which constrain physical
conditions with unprecedented precision.  A UV spectrograph with good
sensitivity down to observed wavelengths of 1000 \AA\ can detect these
new species in absorption systems with redshift \zabs\ $\gtrsim 0.3$,
and at these redshifts, the detailed relationships between the
absorbers and nearby galaxies and large-scale environment can be
studied from the ground.  By observing QSOs at $z = 1.0 - 1.5$, {\it
  HST} has started to exploit extreme-UV QSO absorption lines, but
{\it HST} can only reach a small number of these targets.  A future,
more sensitive UV spectrograph could open up this new discovery space.
\end{abstract}

\noindent \textbf{1. QSO Absorption Lines at Wavelengths $<$ 912 \AA }

\begin{figure*}
   \includegraphics[width=18.3cm, angle=0]{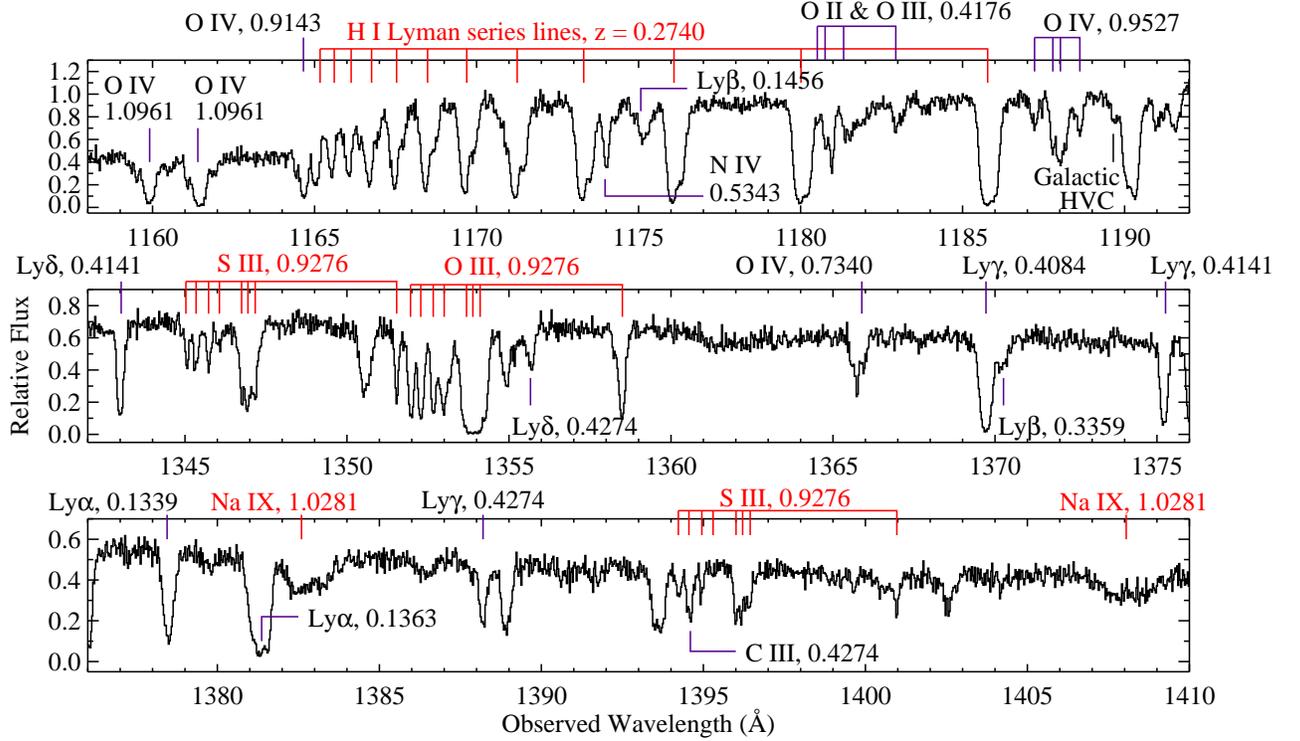}
\caption{\small Examples of the COS spectra obtained from our
  \textit{Hubble Space Telescope} program from the sight lines to
  PG1630+377 ($z_{\rm QSO} = 1.476$, top panel) and PG1206+459
  ($z_{\rm QSO} = 1.163$ middle and lower panels). Various lines of
  metals and H~\textsc{i} are labeled with their redshifts.
  \textbf{Most of the lines in this figure have rest-frame wavelengths
    $< 912$ \AA ; the shortest-wavelength transitions shown here are
    the Na~IX doublet with $\lambda _{\rm rest}$ = 681.7, 694.3
    \AA\ and the O~IV lines at $\lambda _{\rm rest}$ = 553.3, 554.1
    \AA .} Lines and systems of particular interest are indicated in
  red. The high S/N of these data and access to lines down to the
  H~\textsc{i} Lyman limit (and beyond) provide very precise
  constraints on the H~\textsc{i} column densities, the ionization
  state of the gas, the metallicity, gas kinematics, and insights on
  the multiphase physics that governs circumgalactic and intergalactic
  gases. Note that these are only small portions of the spectrum for
  each quasar and are representative of the full
  sample.\label{specsample}}
\end{figure*}

\vspace{-0.2cm} High-resolution ultraviolet spectroscopy provides a unique ability to
study low-density gas/plasma in galaxy disks, halos, and the
intergalactic medium (IGM), i.e., all harbors of present-epoch
baryons. Since stars account for only a small fraction of the baryon
inventory and most of the ordinary matter is in very low-density gases
(Fukugita et al. 1998, ApJ, 503, 518), UV spectroscopy is a crucial
technique for the study of galaxy ecosystems and the cycles of
inflowing and outflowing matter and energy that regulate galaxy
formation. As anticipated by Verner et al. (1994, ApJ, 430, 186), the
deployment of the \textsl{Cosmic Origins Spectrograph} (COS, Green et
al. 2012, ApJ, 744, 60) on the \textit{Hubble Space Telescope} has
demonstrated a particularly powerful new window for UV spectroscopy:
the study of QSO absorption lines in the ``extreme'' ultraviolet (EUV)
at $\lambda <$ 912 \AA . Normally, we assume that EUV absorption lines
cannot be observed because the Galactic ISM prevents observations of
transitions at these wavelengths in the Milky Way.  However, if gas in
a quasar absorption system has a sufficiently high redshift, these
lines are redshifted into the \textit{HST} bandpass; for example, the
Ne~VIII doublet at 770.4,780.3 \AA\ can be studied in QSO absorbers
with redshift $z_{\rm abs} \geq 0.3$ with a spectrograph sensitive
down to 1000 \AA .  Unfortunately, in very high-redshift QSO absorbers
that can be observed from the ground, these EUV lines are ruined by
blending with the thick Ly$\alpha$ forest. However, as illustrated in
Figure~\ref{specsample}, QSOs at $z \approx 1 - 1.5$ are in a ``sweet
spot'' where the EUV lines can be detected but the line density is low
enough so that blending is not severe.  These COS data demonstrate the
potential of EUV lines, but unfortunately, HST+COS can only access a small
  number of these targets in reasonable exposure times.  Moreover,
while the COS spectra have signal-to-noise $\approx 30-50$ per resel,
higher S/N ($\gtrsim 100$) would greatly improve this technique
because the key lines (e.g., Ne~VIII) can be quite weak (see,
e.g. Meiring et al. 2012, arXiv1201.0939).

\begin{figure}
\includegraphics[width=17.5cm, angle=0]{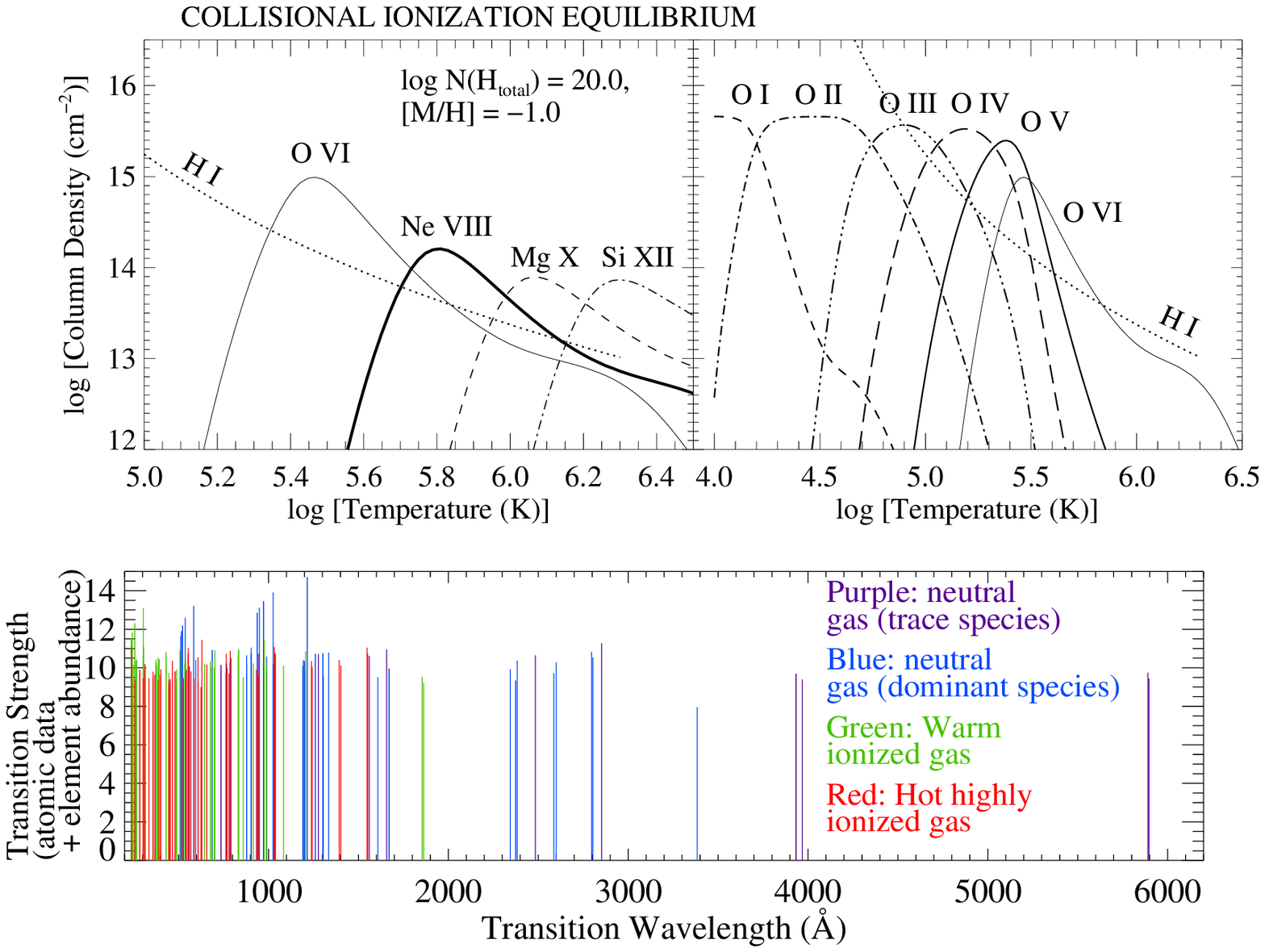}
\caption{\small {\it Upper panels:} Column densities of various metals
  in collisional ionization equilibrium, as a function of temperature,
  for an absorber with $N$(H$_{\rm total}) = 10^{20}$ cm$^{-2}$ and $Z
  = 0.1 Z_{\odot}$. {\it Lower panel:} Strength of resonance lines
  vs. the rest-frame wavelength of the transition from Verner et
  al. (1994, A\&AS, 108, 287) based on the element abundance and
  atomic data (taller lines indicate intrinsically stronger
  transitions). Colors indicate tracers of difference gas phases (see
  legend). In addition to providing access to a larger number of sight
  lines, a future UV spectroscopic facility with greater sensitivity
  could more effectively exploit the rich diagnostics available at
  $\lambda <$ 1000 \AA\ by detecting weaker lines with higher
  signal-to-noise spectra.  \label{euvdemo}}
\end{figure}

Figure~\ref{euvdemo} demonstrates the following unique diagnostics
provided by EUV absorption spectroscopy: First, EUV absorption lines
include species such as Ne~VIII, Mg~X, and Si~XII, and these species
are detectable in plasmas at $T > 10^{6}$ K.  Thus, in the EUV, {\it
  HST} can compete with X-ray telescopes, but {\it HST} has much
better spectral resolution, better sensitivity, and a substantially
larger pool of sufficiently bright targets, which enables more optimal
target selection.  The Astro2010 decadal survey identified the
\textit{International X-ray Observatory} as a top priority for the
next 20 years, and one of the prime science drivers of IXO is the
study of missing baryons and hot gas in low-density gaseous halos and
the IGM using absorption spectroscopy. By using species such as
Ne~VIII and Mg~X, we can pursue this IXO science goal immediately.  We
have already successfully detected Ne~VIII, Mg~X, and other highly
ionized hot-gas tracers (see below).  Second, the EUV includes
transitions of suites of adjacent ions such as O~I, O~II, O~III, O~IV,
O~V, and O~VI or S~II, S~III, S~IV, and S~V (similar sets are
available for C, N, etc). These adjacent ions span a wide range of
temperatures/ionization conditions (see Fig.~\ref{euvdemo}), and
limits on or detections of these species can constrain the physics and
metallicity of QSO absorbers with unprecedented precision [note that
  currently, we typically only have access to scattered ionization
  stages such as O~VI, C~III, and Si~III in low-$N$(H~I) absorbers].
As shown in the lower panel of Fig.~\ref{euvdemo}, the EUV is the
richest region of the spectrum for QSO absorption spectroscopy.
Third, the redshifts of the absorbers are sufficient to bring many H~I
Lyman series lines into the {\it HST} band (see examples in
Figure~\ref{specsample}).  This enables accurate H~I column-density
measurements because the higher Lyman series lines are less likely to
be saturated.  Observations of lower-redshift absorbers often detect
only a few Lyman lines or even only Ly$\alpha$, and often these lines
are badly saturated so $N$(H~I) is highly uncertain. Uncertain
$N$(H~I) measurements lead to uncertain metallicity measurements.
With good constraints on metallicity and physical conditions, key
properties such as mass and mass flow rates can be estimated.

\vspace{-0.2cm}
\noindent \textbf{2. Proof of Concept: First Results from \textit{HST}}

\vspace{-0.2cm}
\noindent \textbf{Galactic Winds Driven by Star Formation and AGN.}
The role of galactic outflows and ``feedback'' is one of the most
pressing issues of current galaxy evolution studies. Some observations
of objects such as Lyman-break galaxies, ULIRGs, and post-starburst
galaxies have revealed dramatic outflows (e.g., Rupke et al. 2005,
ApJS, 160, 87; Tremonti et al. 2007, ApJ, 663, L77; Steidel et
al. 2010, ApJ, 717, 289).  However, since these studies usually use
the ULIRG or the post-starburst galaxy itself as the continuum source,
they suffer from an ambiguity regarding the spatial extent, and hence
the mass, of the outflow.  These investigations also have a limited
suite of diagnostics, e.g., Mg~II or Na~I and nothing else. By using
absorption lines imprinted on background QSOs, these limitations can
be overcome, and the EUV lines turn out to be particularly
interesting.  Multiple examples of different types of outflows are
present in the sample COS data shown in Figure~\ref{specsample}.  For
example, toward PG1206+459 we have clearly detected, at high
significance, a doublet of Na~IX at $z_{\rm abs} = 1.0281$ (see the
lowest panel in Fig.~\ref{specsample}). Na~IX has never been detected
before, but this absorber is also detected in Ne~VIII, Mg~X, Ar~VII,
Ar~VIII, and O~V.

A possibly even more interesting outflow is detected at $z_{\rm abs}$
= 0.9276 in the PG1206+459 spectrum (Tripp et al. 2011, Science, 334,
952).  From Fig.~\ref{specsample} we see that there is a dramatic
cluster of absorption lines at this redshift detected in species such
as S~III and O~III (middle and lower panels). This system is notable
for the following reasons (see Tripp et al. for full details):  First,
we detect the adjacent suites of ions, including O~III, O~IV, O~V, and
O~VI; N~III, N~IV, and N~V; and S~III, S~IV, and S~V.  Second, we
detect Ne~VIII at very high significance.  Third, the Ne~VIII and N~V
velocity centroids are strongly correlated with the centroids of low
ions such as Mg~II, Si~II, and C~II (see Fig.3 in Tripp et
al.). Fourth, while this absorption cluster is clearly a Lyman-limit
absorber with many higher Lyman-series lines, the Lyman limit (LL) is
not black and excellent $N$(H~I) measurements can be obtained.
Finally, there is a post-starburst galaxy with an AGN at an impact
parameter of 68 kpc from the sight line.  These results have
interesting implications: (1) The components in the cluster extend
from $-400$ to +1100 km s$^{-1}$; with these velocities, some
components must be exceeding the escape velocity of the galaxy. (2)
Using the adjacent ions (e.g., SIII/SIV/SV) we can pin down the
ionization state of each component and estimate their total column
densities.  Combined with the large impact parameter (70 kpc) to the
galaxy, this implies that each component carries $\approx 10^{8}$
M$_{\odot}$ of mass in cool, photoionized gas, assuming a standard
thin-shell model (e.g., Tremonti et al. 2007). Other geometries would
give different masses, but an important mass component is implicated
in any case. (3) However, the Ne~VIII and N~V must arise in hot gas
that is correlated with the cool gas -- Ne~VIII/N~V cannot originate
in the cool photoionized gas.  Moreover, this hot gas contains
$10\times$ to $150\times$ more mass than the cool phase.  In addition,
the remarkable correspondence of the Ne~VIII with lower ions suggests
that the outflowing material is also interacting with a hotter
(unseen) phase.  How do species like Mg~II and Si~II survive at
these outflow velocities embedded in such hot gas?

\noindent \textbf{Cold Accretion of Pristine (Low-Metallicity) Gas.}
These spectra also reveal the opposite process: absorbers that are
most naturally explained as cold, {\it inflowing} material, an equally
important topic that is even more poorly constrained by observations.
The partial Lyman limit absorber that produces the Lyman series lines
shown in the top panel of Figure~\ref{specsample} is an example of
apparently infalling, very metal poor gas.  We have analyzed the
metals affiliated with this partial Lyman limit system (Ribaudo et
al. 2011, ApJ, 743, 207), and we find that the logarithmic metallicity
is only [Mg/H] = $-1.71 \pm 0.06$.  Moreover, we have
spectroscopically identified and studied a nearby galaxy at the
redshift of the Lyman limit absorber at an impact parameter of 37 kpc.
Interestingly, that galaxy has a metallicity that is almost two orders
of magnitude higher, [O/H]$_{\rm galaxy} -0.20 \pm 0.15$.  This
absorber may represent nearly primordial material that is accreting
onto the galaxy via cold-mode accretion (Kere\v{s} et al. 2005), but
other explanations remain viable. Subsequently, we studied all LL
absorbers (16.0 $<$ log $N$(H~I) $<$ 19.) in our data combined with
measurements from the \textit{HST} archive and literature (Lehner et
al. 2012, in prep.), and we find that 50\% of LL have very low-metallicity
($Z \leq 0.03 Z_{\odot}$). Our survey has tripled the sample of LL absorbers
with good metallicity measurements at $z < 1$, but the sample is still
small (28 systems total).

\noindent \textbf{Requirements for a Future UV Telescope.} Technical
concepts are deferred to the second RFI, but the key technical
requirements for this science can be briefly summarized. While
\textit{HST} has begun to observe QSO absorption lines at $\lambda
_{\rm rest} < 912$ \AA , the number of $z_{\rm QSO} = 1 - 2$ QSOs
bright enough for \textit{HST} is extremely small.  To exploit this
discovery space, a future UV spectrograph must have substantially
better sensitivity than \textit{HST}+COS, good spectral resolution
(comparable to STIS and COS), and wavelength coverage down to at least
1150 \AA\ and preferrably down to $\approx$ 1000 \AA .


\end{document}